\documentclass[aps, prl,showpacs,twocolumn,preprintnumbers,
amsmath,amssymb]{revtex4}

\usepackage{graphicx}
\usepackage{dcolumn}
\usepackage{bm}
\usepackage{ifthen}

\begin{document}

\title{Ferroelectricity driven by 
the non-centrosymmetric magnetic ordering
in multiferroic TbMn$_2$O$_5$:
a first-principles study}
\author{Chenjie Wang}
\affiliation{Key Laboratory of Quantum Information,
University of Science and Technology
of China, Hefei, 230026, People's Republic of China}
\author{Guang-Can Guo}
\affiliation{Key Laboratory of Quantum Information,
University of Science and Technology
of China, Hefei, 230026, People's Republic of China}

\author{Lixin He\footnote{corresponding author, 
Email address: helx@ustc.edu.cn} 
}
\affiliation{Key Laboratory of Quantum Information,
University of Science and Technology
of China, Hefei, 230026, People's Republic of China}
\date{\today}

\begin{abstract}
The ground state structural, electronic and magnetic
properties of multiferroic TbMn$_2$O$_5$ are investigated via
first-principles calculations. We 
show that the ferroelectricity in TbMn$_2$O$_5$ is 
driven by the non-centrosymmetric magnetic ordering,
without invoking the spin-orbit coupling and non-collinear spins. 
The {\it intrinsic} electric polarization in this compound is
calculated to be 1187 $nC\cdot$ cm$^{-2}$, 
an order of magnitude larger than previously thought. 
\end{abstract}


\maketitle


Multiferroics with magnetic and electric ordering united
in a single phase were thought to be 
rare\cite{fiebig05,hill05}.
Surprisingly, a large class of manganese
oxides (RMnO$_3$ \cite{kimura03,goto04}, 
and RMn$_2$O$_5$ \cite{hur04,chapon04,blake05} 
with R=Y, Tb, Dy, etc.) has recently been discovered to
be multiferroic.
Unlike the traditional multiferroics where the two order parameters only
couple weakly\cite{fiebig05},  
the newly discovered materials possess strong magnetoelectric (ME)
coupling, resulting in various novel physical effects.
One of the most prominent examples is given by
TbMn$_2$O$_5$\cite{hur04,cheong07}, 
which displays clear
correlated anomalies of the dielectric
constant $\epsilon$ with the spin ordering \cite{hur04}.
More strikingly, the electric polarization in this material
can be reversed by applying a magnetic field \cite{hur04}.
The remarkable ME effects revealed 
in these materials have attracted great attention 
\cite{fiebig05,kagomiya03,kimura03,goto04,blake05,hur04, 
chapon04,aguilar06,cheong07,katsura05, sergienko06}   
because of the fascinating physics
and their potential applications in novel multifunctional 
ME devices. 

Although great effort has been devoted to
understanding the fundamental mechanism of the
giant ME coupling,
our knowledge of the 
manganese oxide multiferroics 
is still very limited and full of puzzles.
For example, experimental data show that 
the structure of TbMn$_2$O$_5$
has space group \emph{Pbam} \cite{alonso97}, which
includes spatial inversion ($R^{-1}$) symmetry.
It is therefore puzzling that the compound develops
spontaneous electric polarizations.
It has been suspected \cite{kagomiya03, chapon04}
that the actual symmetry group of TbMn$_2$O$_5$
is \emph{Pb}2$_1$\emph{m}, allowing polarizations.
Indeed, there are several experiments supporting this hypothesis
\cite{chapon04,blake05,aguilar06}.
Nevertheless, no {\it direct}
evidence of the lower symmetry has yet been found \cite{chapon04, blake05}.
Theoretically, the origin of the giant ME coupling and
the ferroelectricity in these materials
is still under intensive debates
\cite{chapon04, cheong07,katsura05, sergienko06}.
One of the fundamental questions that remain unsolved is 
whether the spin-orbit interaction \cite{katsura05, sergienko06}  
is essential for the ME coupling and ferroelectricity in these materials.

In this letter, we carry out a first-principles study of TbMn$_2$O$_5$,
to identify its crystal structure and 
clarify the microscopic origin of the ferroelectricity.
We compute physical quantities that can be directly compared to the
experiments.
To the best of our knowledge, no such 
study has yet been done for TbMn$_2$O$_5$ (and other 
RMn$_2$O$_5$ compounds \cite{blake05}), 
because it has very complicate incommensurate
anti-ferromagnetic (AFM) structure 
with the propagation vector ${\bf k} \approx (0.48, 0, 0.32)$.
To accommodate the magnetic structure, one needs a
huge supercell, which is computationally prohibitive.
Instead, we use a 2$\times$1$\times$1 supercell,
equivalent to approximating the propagation vector ${\bf k}=(0.5, 0, 0)$.
The validity of this approximation will be
justified later in the text.
Our results show that
the ferroelectricity in TbMn$_2$O$_5$ is
driven by the non-centrosymmetric magnetic ordering and the asymmetric exchange
interactions, 
without invoking the spin-orbit coupling and non-collinear spins. 
The {\it intrinsic} electric polarization in this compound is
calculated to be 1187 $nC\cdot$ cm$^{-2}$, 
much larger than previously thought for this type of compound
\cite{fiebig05,hur04}.

\begin{table}
\caption{Comparison of the calculated and measured
structural parameters of TbMn$_2$O$_5$. The lattice constants
are given in \AA. Atoms that occupy the same Wyckoff positions
are shown only once.}
\begin{center}
\begin{tabular}{l ccc ccc}
\hline \hline & \multicolumn{3}{c}{Theory (\emph{Pb}2$_1$\emph{m)}}
&\multicolumn{3}{c}{Experiment
(\emph{Pbam})} \\
                 &$a$ &$b$ &$c$ &$a$ &$b$ &$c$ \\\hline
$a$, $b$, $c$  & 7.3014 & 8.5393 & 5.6056 &7.3251 & 8.5168 & 5.6750  \\
\hline
Tb$^{3+}$& 0.1410  & 0.1733 & 0        &  0.1399 &  0.1726  & 0      \\
         & 0.6404  & 0.3270 & 0                                      \\
Mn$^{4+}$& 0.0001  & 0.5003 & 0.2558   &  0      &  0.5     & 0.2618 \\
Mn$^{3+}$& 0.4012  & 0.3558 & 0.5      &  0.4120 &  0.3510  & 0.5    \\
         & 0.9016  & 0.1456 & 0.5                                    \\
O$_1$    & 0.0008  & 0.0002 & 0.2709   &  0      &  0       & 0.2710 \\
O$_2$    & 0.1645  & 0.4480 & 0        &  0.1617 &  0.4463  & 0      \\
         & 0.6648  & 0.0517 & 0                                      \\
O$_3$    & 0.1560  & 0.4329 & 0.5      &  0.1528 &  0.4324  & 0.5    \\
         & 0.6571  & 0.0655 & 0.5                                    \\
O$_4$    & 0.3977  & 0.2077 & 0.2438   &  0.3973 &  0.2062  & 0.2483 \\
         & 0.8959  & 0.2919 & 0.7579                                 \\
\hline
\end{tabular}
\label{tab:atom}
\end{center}
\end{table}

The calculations are based on the density functional theory (DFT) within
the spin-polarized generalized gradient approximation (GGA)
\cite{perdew96} implemented
in the Vienna \emph{ab initio} Simulations Package (VASP)
\cite{kresse93,kresse96}. 
The projector
augmented-wave (PAW) pseudopotentials \cite{blochl94} with a 500 eV plane-wave
cutoff are used.
A $1\times2\times4$ Monkhorst-Pack k-points mesh converges 
very well the results.
We use the
collinear spin approximation without the spin-orbit coupling.
Our results agree very well with the known experiments,
indicating that these approximations capture the essential physics in
TbMn$_2$O$_5$.

\begin{figure}
\centering
\includegraphics[width=3.5in]{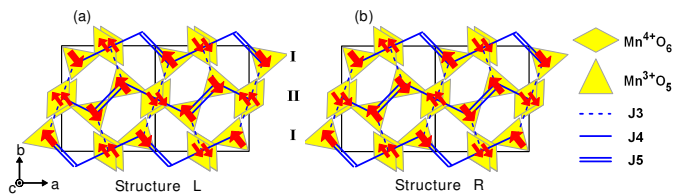}
\caption{(Color online) The ground state spin configurations
for two energetically degenerate structure $L$ and $R$.
The diamonds and triangles denote the Mn$^{4+}$O$_6$  octahedra
and Mn$^{3+}$O$_5$ pyramids respectively.
The dashed, single and double lines represent
$J_3$, $J_4$ and $J_5$ exchange interactions respectively, following
Ref. \onlinecite{chapon04}.
}
\label{fig:magneticstr}
\end{figure}

The crystal structure of TbMn$_2$O$_5$ is orthorhombic,
with four TbMn$_2$O$_5$ formula units 
(32 atoms in total) per primitive cell,
containing  Mn$^{4+}$O$_6$ octahedra 
and Mn$^{3+}$O$_5$ pyramids \cite{alonso97}.
We relax the crystal structure beginning with
the experimental structural parameters\cite{alonso97}, 
listed in Table~\ref{tab:atom}.
The most stable structure we found
has the spin configuration identical to that was proposed
in Ref. \onlinecite{chapon04}, as illustrated in
Fig. \ref{fig:magneticstr}(a).
It also has an energetically degenerate structure,
shown in Fig. \ref{fig:magneticstr}(b). We denote the two structures
``left'' ($L$) and ``right'' ($R$)  respectively.
In these magnetic structures, Mn$^{4+}$ form an 
AFM square lattice in the $ab$ plane, whereas Mn$^{3+}$ couples to
Mn$^{4+}$ either antiferromagnetically via $J_4$ along $a$ axis or
with alternating sign via $J_3$ along $b$ axis.
Mn$^{3+}$ ions in two connected pyramids also couple
antiferromagnetically through $J_5$. Here, we adopt the notations
$J_3$, $J_4$ and $J_5$ from Ref. \onlinecite{chapon04}, and define
the $J_3$ to be the Mn$^{4+}$- Mn$^{3+}$ superexchange interaction
through pyramidal base corners, and $J_4$ the superexchange
interaction through the pyramidal apex, as indicated in Fig.
\ref{fig:magneticstr}. We label the two different Mn$^{4+}$ chains
along the $a$ axis I, II respectively, also following Ref.
\onlinecite{chapon04}.
The magnetic structure of $R$ can be obtained from $L$ by shifting
chain II to the right (or to the left) by one unit cell along the
$a$ axis \cite{chapon04}. 

The calculated structural parameters for the structure $L$
are listed in
Table~\ref{tab:atom}, comparing with the experimental data,
whereas the structure $R$ is a mirror image of $L$ about $ac$-plane.
The calculated structural parameters are in extremely good agreement
with the experimental data.
The errors of lattice constants are about 1\%, typical errors for GGA.
The atom positions are also extremely close to what was obtained experimentally.
However, the small atomic displacements 
lower the structural symmetry to the long searched
\emph{Pb}2$_1$\emph{m} polar group.
To see how this happen,
we take Mn$^{3+}$ ions as an example.
In the \emph{Pbam} structure, Mn$^{3+}$ has one Wyckoff position (h in Wyckoff
notation) that has four equivalent sites shown in
Table \ref{tab:Mn3+}. 
However, in the \emph{Pb}2$_1$\emph{m} structure,
it splits into two Wyckoff positions b(1), b(2), each
has two equivalent sites. 
$\delta x$ and $\delta y$ in Table \ref{tab:Mn3+} 
denote the atomic displacements from the high symmetry positions
along the $a$ and $b$ axes respectively. 
The displacements along the $a$ axis are of mirror symmetry, whereas the
displacements along the $b$ axis are not, allowing polarizations. The
atomic displacements can be easily calculated from Table
~\ref{tab:atom}. We see that the displacements are extremely small,
usually are of the order of $\sim$ 10$^{-4}$ of the lattice
constants and the largest atom displacements come from Mn$^{3+}$,
$\delta y$ $\sim$ 10$^{-3}$ of the lattice constants. 
Therefore the low symmetry structure can not be directly
determined experimentally, and only the anomalies of 
the atomic displacement parameters (ADPs)
were observed \cite{chapon04}.
We also artifically construct a high symmetry structure
by symmetrizing structure L and R 
according to the \emph{Pbam} symmetry,     
which we refer as
structure $H$ in the following discussions.

\begin{table}
\begin{center}
\caption{Comparison of the Mn$^{3+}$ positions in \emph{Pbam} symmetry
and \emph{Pb}2$_1$\emph{m} symmetry.}
\begin{tabular}{cccc}
\hline \hline
\multicolumn{2}{c}{\emph{Pbam}} &\multicolumn{2}{c}{\emph{Pb}2$_1$\emph{m}} \\
\hline
h &(x, y, 1/2)          & b(1) & ( x, y, 1/2)         \\
h &(-x+1/2, y+1/2, 1/2) & b(1) & (-x+1/2, y+1/2, 1/2)\\
h &(x+1/2, -y+1/2, 1/2) & b(2) & (x+1/2+$\delta {x}$, -y+1/2-$\delta {y}$, 1/2)\\
h &(-x, -y, 1/2)        & b(2) & (-x-$\delta {x}$,    -y-$\delta {y}$, 1/2)       \\
\hline
\end{tabular}
\label{tab:Mn3+}
\end{center}
\end{table}

\begin{figure}
\centering
\includegraphics[width=3.2in]{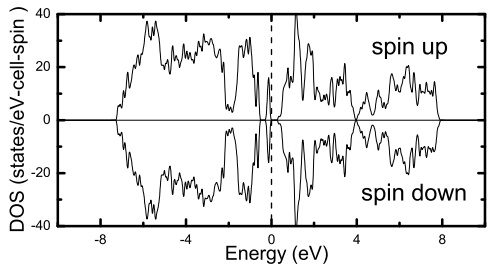}
\caption{The total DOS for TbMn$_2$O$_5$ of structures $L$ and $R$.
The dashed line indicates valence-band maximum.}
\label{fig:DOS}
\end{figure}

The calculated density of states (DOS) of
structure $L$ (and also $R$)
is shown in Fig.\ref{fig:DOS}.
The DOS for spin-up and spin-down electrons is identical
as expected for an AFM state.
The DOS exhibits a small but clear band gap (about 0.4 eV), confirming the
experimental fact that TbMn$_2$O$_5$ is an insulator.
However, it is well known that GGA greatly underestimates the band gap,
especially for the 3d compounds.
The local magnetic moments are estimated for Mn$^{3+}$ to be $\sim$
2.37 $\mu_B$, and for Mn$^{4+}$ to be $\sim$ 1.64 $\mu_B$, in good
agreement with the refined magnetic moments \cite{chapon04}. 

One of the strong evidences suggesting that the low temperature
structure of TbMn$_2$O$_5$ has space group \emph{Pb}2$_1$\emph{m} is that
some Raman active modes become also IR active
\cite{aguilar06} at low temperature,
which is forbidden by higher \emph{Pbam} symmetry.
To clarify this problem, we analyze the zone-center phonons.
The symmetry analyses are performed on the 32-atom primitive
cell \cite{he02}.
For the high symmetry structure (\emph{Pbam}),
the total 96 modes, 
are decomposed into 8 irreducible representations
(irreps):
%
\begin{equation} 
\Gamma=B_{1u}\oplus B_{2u}\oplus B_{3u}\oplus A_{g} 
\oplus B_{1g}\oplus B_{2g}\oplus B_{3g}\oplus A_{u}\,.
\label{eq:high-symm-modes}
\end{equation}
Among them $B_{3u}$, $B_{2u}$ and $B_{1u}$ modes are IR active,
polarized along the \emph{a}, \emph{b} and \emph{c} axes, respectively.
$B_{1g}$, $B_{2g}$, $B_{3g}$ and $A_{g}$ modes are Raman active while
$A_{u}$ modes are silent. 
As we see, the Raman- and IR-active modes do not couple.
However, the real crystal structure possesses
a \emph{Pb}2$_1$\emph{m} symmetry,
whose 96 phonons can be decomposed into 4 irreps:
\begin{equation}
\Gamma=A_1\oplus B_1\oplus B_2\oplus A_2 \, .
\label{eq:low-symm-modes}
\end{equation}
We found all modes are Raman active,
among them $A_1$, $B_1$ and $B_2$ are also IR active,
and are polarized along the $b$, $c$, and $a$ axes
respectively \cite{aguilar06}.
Detailed analyses show that the $A_1$ modes are coupled from the $A_{g}$
and $B_{2u}$ modes. 
We then calculate the phonon frequencies via a frozen-phonon technique
\cite{he02}, as well as the oscillator strengths for the IR modes 
\cite{unpublished}.
The calculated phonon frequencies are all in good agreement
with the experiments, generally within 20 cm$^{-1}$
from the experimental values.
All phonons are found to be stable, i.e., no soft phonon
has been found as in traditional ferroelectrics.
In this letter, we show only the $b$-axis-polar $A_1$ phonons in
Table \ref{tab:B2uphonon}, with frequencies and their oscillator
strengths. To make a good contact with experiments, we divide the
phonons into two presentations $A_g$ and $B_{2u}$ according to their
major symmetry character. The experimental frequencies of the $A_g$
phonons are extracted from Raman spectra \cite{mihailova05}, while
the $B_{2u}$ phonon frequencies are extracted from IR spectra
\cite{aguilar06}. 
The overall calculated oscillator strengths of $A_1$ modes 
are in good agreement with the experimental values
\cite{aguilar06}.
More importantly, the 693 cm$^{-1}$ $A_g$ Raman mode, that was
found also IR active with oscillator
strength $S_{\lambda}$=0.001 in the experiment \cite{aguilar06},
is well reproduced in the calculations,
with oscillator strength  $S_{\lambda}$=0.004, 
therefore, confirming that
the ground state structure is indeed of \emph{Pb}2$_1$\emph{m}
symmetry.

We next calculate the spontaneous
polarization in the compound using the Berry-phase theory of
polarization implemented in VASP \cite{king-smith93}. 
The {\it intrinsic} polarization in this material is
calculated to be 1187 $nC\cdot$cm$^{-2}$ along the $b$ axis. 
This value is an order of magnitude smaller than 
that of the traditional ferroelectrics. e.g., BaTiO$_3$, yet it is
about 30 times larger than the {\it currently} measured experimental value
($\sim$ 40 $nC\cdot$cm$^{-2}$ \cite{hur04}) for this compound.
The large discrepancy between calculated and experimental polarizations
might come from the approximations we used
in the calculations.
For example, we approximate the magnetic propagation vector 
$k_z$=0.32 by zero. We also
ignore the spin-orbit coupling and assuming collinear spins. Without these
approximations, the polarization might be smaller.
On the other hand, the experiment \cite{hur04} measured
polycrystalline samples,
in which grains polarize along different directions,
canceling each other,
therefore might greatly underestimated the {\it intrinsic} polarization.
We believe a high quality single crystal sample should
enhance the measured electric polarization.

\begin{table}
\tabcolsep 1.8mm 
\caption{Calculated phonon frequencies ($\omega$)
and oscillator strengths ($S_{\lambda}$) of IR-active A$_1$
modes compared with experimental values. The modes are divided into B$_{2u}$
and A$_{g}$ representations according to their major symmetry
character. The experimental values of B$_{2u}$ modes are taken from
Ref. \onlinecite{aguilar06}, whereas those of A$_{g}$ modes are
taken from Ref. \onlinecite{mihailova05}. }
\begin{tabular}{cccccccc}
\hline \hline
\multicolumn{4}{c}{B$_{2u}$} & \multicolumn{4}{c}{A$_{g}$}\\
\multicolumn{2}{c}{GGA}   & \multicolumn{2}{c}{Exper.}  & \multicolumn{2}{c}{GGA}  &\multicolumn{2}{c}{Exper.}\\
$\omega$ & $S_{\lambda}$  &$\omega$ &  $S_{\lambda}$  &$\omega$ &  $S_{\lambda}$  &$\omega$ &  $S_{\lambda}$  \\
\hline
100.7 & 0.11 & 97.2   & 0.42 & 110.1 & 0.005&      & \\
158.0 & 0.44 & 168.9  & 0.46 & 136.9 & $\sim$0 &      &\\
162.8 & 0.50 & 171.9  & 0.30 & 221.6 & 0.009  & 215  &\\
224.8 & 0.24 & 222.2  & 0.11 & 235.1 & 0.011  & 221  &\\
267.3 & 0.12 & 256.8  & 0.17 & 312.5 & 0.022  & 334  &\\
316.7 & 0.82 & 333.4  & 0.17 & 340.2 & 0.011  & 350  &\\
351.3 & 0.09 & 386    & 0.02 & 405.6 & 0.0006 & 412  &\\
412.5 & 0.13 & 422.3  & 0.28 & 445.1 & 0.125  &      &\\
439.5 & 4.81 & 453.2  & 3.43 & 489.2 & 0.013  & 500  &\\
471.0 & 1.23 & 481.8  & 2.86 & 529.2 & 0.005  & 537  &\\
533.5 & 0.27 & 538.2  & 0.25 & 612.0 & $\sim$0 & 621  &\\
549.3 & 0.12 & 567.3  & 0.52 & 613.5 & 0.0003 & 631  &\\
625.0 & 0.36 & 636.6  & 0.27 & 673.6 & 0.004&
693\footnote{experimentally observed IR active and 703 cm$^{-1}$
measured by
Aguilar et al.\cite{aguilar06}} & 0.001\\
667.2 & 0.009 & 688.3  & 0.003\\
\hline
\end{tabular}
\label{tab:B2uphonon}
\end{table}

To further elucidate the origin of the polarization,
we also calculate the spontaneous polarization for the high symmetry
structure $H$ and get ${\bf P}$= 228 $nC\cdot$cm$^{-2}$.
In this case, the crystal structure has $R^{-1}$ symmetry, 
however, because the special spin configuration 
(see Fig. \ref{fig:magneticstr}) 
of the Mn$^{3+}$-Mn$^{4+}$-Mn$^{3+}$ chains 
along the $b$ axis 
breaks the $R^{-1}$ and the combined spatial inversion
and time reversal  
[($RT)^{-1}$] symmetry of 
the magnetic structure, 
the electron wavefunctions have lower symmetry than the lattice,
resulting in polarization. 
The {\it electronic} symmetry breaking will further couple to the 
lattice and lead to lattice distortion. 
Therefore, when holding atoms fixed at the centrosymmetric
structure, turning on the magnetic order does two things: it generates
a purely electronic polarization of 228 $nC\cdot$cm$^{-2}$, and
it also applies forces to the atoms.  
These forces give rise to atomic displacements that yield
an additional 959 $nC\cdot$cm$^{-2}$.

\begin{figure}
\centering
\includegraphics[width=3.0in]{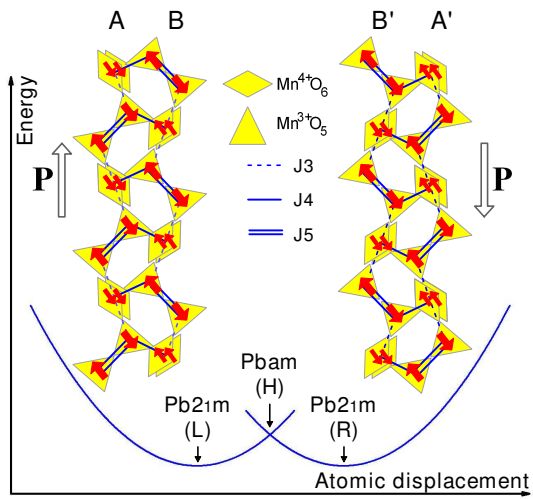}
\caption{ (Color online) An asymmetric-spin-chain model showing the
sketch of energy surfaces of
structure $L$ and $R$ vs. the atomic displacements
from the high symmetry structure.
The diamonds and triangles denote the Mn$^{4+}$O$_6$  octahedra
and Mn$^{3+}$O$_5$ pyramids respectively.
  }
\label{fig:chain-model}
\end{figure}

We now discuss the microscopic mechanism of the coupling between magnetic 
order and lattice.
Figure \ref{fig:chain-model} depicts the energy surfaces of the
structure $L$ and $R$ vs. atomic displacements.
The two structures degenerate in energy at the
high symmetry structure $H$ (\emph{Pbam}).
The magnetic structures of $L$ and $R$ can be simplified as
Mn$^{3+}$-Mn$^{4+}$-Mn$^{3+}$ segments linked along
the  $b$ axis via $J_5$.
Inside the segments, Mn$^{3+}$
interact with Mn$^{4+}$ via exchange interaction $J_3$.
In each segment, the Mn$^{3+}$ ions above and below Mn$^{4+}$ ions
have opposite spins. For structure $L$,
Mn$^{4+}$ always has the {\it same} sign of spin with the lower Mn$^{3+}$,
but {\it opposite} sign with the
upper one in each segment, whereas the
opposite is true for structure $R$. In the high
symmetry structure $H$, 
the Mn$^{3+}$ and Mn$^{4+}$ ions with 
the same sign of spin could move closer to each other
to minimize the exchange energies\cite{cheong07}.
The formal description of the coupling between the spin
and the lattice is given by  \cite{fiebig05,unpublished}
\begin{equation}
-\sum_{ij\lambda}{\partial J_{ij} \over \partial
u_{\lambda}}u_{\lambda}{\bf S}_i\cdot{\bf S}_j \, ,
\end{equation}
where $J_{ij}$ is the
exchange interaction between the magnetic moments
${\bf S}_i$, ${\bf S}_j$ of the $i$-th and $j$-th atoms, 
and ${\bf u}_{\lambda}$ is the $\lambda$-th zone-center
phonon calculated for structure $H$. 
This term is not zero, provides the magnetic 
structure (spin configuration)
$\{ {\bf S}_i \}$ does not have the $R^{-1}$ and the
$(RT)^{-1}$ symmetry, as e.g., in TbMn$_2$O$_5$. 
Therefore the high symmetry structure $H$ is an
unstable point on the energy surface as illustrated 
in Fig.~\ref{fig:chain-model}.  
The structure will spontaneously relax to
$L$ or $R$ according to the spin configuration at the high symmetry
point.
Furthermore, as long as the asymmetric spin structures of the
Mn$^{3+}$-Mn$^{4+}$-Mn$^{3+}$ chains are persevered, the propagation
vector k$_z$ will not change the essential physics 
in this system \cite{radaelli06} 
which justify our approximation of setting
$k_z$=0.

To conclude, we have shown via a first-principle study
on TbMn$_2$O$_5$, how the ferroelectricity is 
driven by the non-centrosymmetric magnetic ordering,
without invoking the spin-orbit coupling and non-collinear spins.
We believe that this work sheds new light 
on the fundamental mechanism of the giant magnetoelectric 
coupling in the multiferroics, especially for the
manganese oxides. 
Surprisingly, the calculated {\it intrinsic} polarization 
in TbMn$_2$O$_5$ is as large as 1187 $nC\cdot$cm$^{-2}$, 
much larger than previously thought for this type of compound.
The mechanism revealed here
also provides useful
guidance for searching for novel magnetoelectric materials.

L.H. would like to thank D. Vanderbilt for valuable suggestions. This work
was supported by the Chinese National Fundamental
Research Program 2006CB921900, 
the Innovation funds and ``Hundreds of Talents''
program from Chinese Academy of Sciences.




\end{document}